\documentclass[conference]{IEEEtran}

\usepackage[dvips]{graphicx}
\usepackage{amsmath,amssymb}
\usepackage{algorithm}
\usepackage{algorithmic}

\usepackage{color}
\usepackage{framed}

\newcommand{\specialcell}[2][c]{%
	\begin{tabular}[#1]{@{}c@{}}#2\end{tabular}}

\newcounter{t0d0_counter}
\newcommand{\nofixme}[1]{
}
\newcommand{\fixme}[1]{
  \stepcounter{t0d0_counter}
  \definecolor{shadecolor}{rgb}{1,1,0} 
  \begin{shaded}
  T0D0 \arabic{t0d0_counter}: #1
  \end{shaded}
}

\usepackage{graphicx}
\usepackage{hyperref}
\usepackage{comment}
\usepackage{balance}

\usepackage{array}
\newcolumntype{H}{>{\setbox0=\hbox\bgroup}c<{\egroup}@{}}

\usepackage{amsmath}
\usepackage{cite}
\usepackage{multirow}  
\usepackage{booktabs}

\usepackage{tablefootnote}

\usepackage{xcolor}

\IEEEoverridecommandlockouts

\begin{document}

\title{Feasibility Study on CCTV-aware Routing and Navigation for Privacy, Anonymity, and Safety. \\
{\large Jyv\"askyl\"a -- Case-study of the First City to Benefit from CCTV-aware Technology.}
}

\date{}

\author{
\IEEEauthorblockN{
Tuomo Lahtinen\IEEEauthorrefmark{4},
Lauri Sintonen\IEEEauthorrefmark{4},
Hannu Turtiainen\IEEEauthorrefmark{5},
Andrei Costin\IEEEauthorrefmark{5} \textsuperscript{$\circledast$}
}
\IEEEauthorblockA{
\textit{University of Jyv\"askyl\"a}\\
Jyv\"askyl\"a, Finland \\
\IEEEauthorrefmark{5}\{turthzu,ancostin\}@jyu.fi, \\
\IEEEauthorrefmark{4}\{tuomo.t.lahtinen,lauri.m.j.sintonen\}@student.jyu.fi
}
\thanks{$\circledast$ Corresponding and original idea's author.}
\thanks{This paper is best viewed in color, but has been converted to grayscale according to paper submission instructions.}
}

\maketitle


\begin{abstract}

In order to withstand the ever-increasing invasion of privacy by 
CCTV cameras and technologies, on par \emph{CCTV-aware solutions} must 
exist that provide privacy, safety, and cybersecurity features. 
We argue that a first important step towards such CCTV-aware solutions 
must be a mapping system (e.g., Google Maps, OpenStreetMap) that provides 
both privacy and safety routing and navigation options. 
Unfortunately, to the best of our knowledge, there are no mapping nor 
navigation systems that support CCTV-privacy and CCTV-safety routing options. 

At the same time, in order to move the privacy vs. safety debate related to 
CCTV surveillance cameras from purely subjective to data-driven and 
evidence-based domain, it would require to implement and evaluate a 
globally-relevant CCTV-aware routing and navigation system. 
This however, would require a tremendous initial effort without any guarantees 
about the feasibility and the end result.
Therefore, we propose first a preliminary evaluation of such CCTV-aware 
technology at a small scale, for example a medium-sized European city.

In this paper, we explore the feasibility of a CCTV-aware routing and 
navigation solution. The aim of this feasibility exploration is to 
understand what are the main impacts of CCTV on privacy, and what are 
the challenges and benefits to building such technology.
We evaluate our approach on seven (7) pedestrian walking routes within the 
downtown area of the city of Jyv\"askyl\"a, Finland. We first map a total of 
450 CCTV cameras, and then experiment with routing and navigation under 
several different configurations to \emph{coarsely model} the possible cameras' 
parameters and coverage from the real-world. 

We report two main results. 
First, our preliminary findings support the overall feasibility of our approach. 
Second, the results also reveal a data-driven worrying reality for persons 
wishing to preserve their privacy/anonymity as their main living choice. 
When modelling cameras at their low performance end, a privacy-preserving 
route has on average a 1.5x distance increase when compared to generic routing. 
When modelling cameras at their medium-to-high performance end, 
a privacy-preserving route has on average a 5.0x distance increase, 
while in some cases there are no routes possible at all, i.e., 
privacy/anonymity must be given up to reach the destination. 
These results further support and encourage both global mapping of CCTV cameras 
and refinements to camera modelling and underlying technology in the context of 
more accurate CCTV-aware routing and navigation. 

\end{abstract}


\section{Introduction}
\label{sec:intro}

CCTV camera surveillance has been spread all over the world and major 
cities have a very dense network of surveillance cameras. People can not move 
in streets or cities without getting caught by surveillance.
Currently, it is estimated there are about 770 millions CCTV cameras 
around the globe, and their number is casually predicted to surpass 
1 billion in 2021~\cite{cnbc2019billion}. 
For example, a study in UK finds that on average a person enters 
a CCTV camera view 300 times a day~\cite{caughtoncamera2019CCTVlond}.
A similar study in the US puts that number at 50+ times a day, despite the 
more worrying fact that the average US respondent \emph{assumed 
it was 4 cameras or less}~\cite{ipvm2016us}.
This patronizing way of measuring people every movement is frightening when 
it is combined with face recognition. Hu et al.~\cite{hu2004survey} surveyed 
and described visual surveillance and great progress in face recognition. 
Since then face recognition has been developed a lot and very accurate 
surveillance is possible with modern cameras.

In order to withstand this ever-increasing invasion of privacy by 
CCTV cameras and technologies, on par \emph{CCTV-aware solutions} must 
exist that provide privacy, safety, and cybersecurity features. 
We argue that a first important step towards such CCTV-aware solutions 
must be a mapping system (e.g., Google Maps, OpenStreetMap) that provides 
both privacy and safety routing and navigation options.

At present, there is a myriad of route planning algorithms, software, and 
services~\cite{osm-route-online,osm-route-offline,luxen2011real,delling2009engineering,bast2016route,szczerba2000robust}.
Therefore a wide variety of solutions are available: open- and closed-source, 
online and offline, free and paid. 
However, to the best of our knowledge, none of the currently available 
algorithms, software, and services provide \emph{CCTV-aware} route planning and navigation.
In our approach, we propose two routing options, namely 
\emph{privacy-first} and \emph{safety-first}. 
Privacy-first routing calculates and optimizes the route in order to avoid 
the field of view of cameras along the generated route between departure 
and arrival points.
Such a scenario is desirable anytime (e.g., day-time, night-time) when 
privacy/anonymity is important, or there are no GDPR-enforceable 
guarantees on the CCTV footage.
Safety-first routing is somewhat opposite to the goals of privacy-first routing. 
It calculates and optimizes the route in order to guarantee the user that the 
route follows the streets with as many cameras as possible.
We call it safety-first because the intention is to increase the chances of 
physical safety for the user.
Such a scenario is desirable for example when visiting known unsafe 
or unknown areas, or when navigating at night.
Generally speaking, in a model where all the CCTV cameras are mapped, 
the \emph{safety-first} would be very close to normal routing because 
we expect the cameras to be everywhere. However, there may still be cases 
where \emph{safety-first} routes would be different from normal routes 
because route optimization does not take time nor distance into account 
but rather maximizing the exposure to the CCTV cameras.
However it is important to note that at present none of these two routing 
modes can provide 100\% guarantee. One reason is the lack of accurate and 
up-to-date data about CCTV camera properties (e.g., model, position, owner). 
Another reason is that our model is not yet fully refined and does not support 
sub-meter routing accuracy. 

To the best of our knowledge, we are the first to propose and work on such 
features at present and we are unaware of any project or service developing 
or offering such route planning options.
We build our current work atop and in relation to another contribution of ours, 
where we developed and demonstrated the first and only Computer Vision (CV) 
models to scalably and accurately detect CCTV camera object in images 
(e.g., street-level, indoors) and we achieve an accuracy of up to 98.7\%~\cite{turtiainen2020towards}.
Our bigger vision is that using those CV models we can instantly, accurately 
and globally map the vast majority of CCTV cameras visible on street-level 
imagery services such as Google Street View~\cite{googlesv}, 
Yandex Street Panoramas, Baidu Total View, Mapillary. 
In our vision, once this localization and mapping is completed, we can 
experiment further at large scale with the CCTV-aware routing and navigation 
technology introduced in current paper. 


\subsection{Contributions}
\label{sec:contrib}

\begin{itemize}
\itemsep0em

\item We are the first to introduce and model CCTV-aware \emph{privacy-first} and 
\emph{safety-first} routing scenarios as relevant for modern digitized lifestyle. 

\item We model, quantify and report CCTV cameras' impact on privacy for 
pedestrian routes using a modern city downtown area as experimental ground.

\item We outline the challenges and the benefits of pursuing these 
experiments and models at global scale.

\end{itemize}


\section{Related Work}
\label{sec:related}

There are hundreds of OSM-based services and routers
~\cite{osm-route-online,osm-route-offline,osm-based,luxen2011real,bast2016route,szczerba2000robust}.
Many of the routers are related to basic driving, walking or cycling
navigation, and others focus on more specific use-cases ranging from highlighting
curvy roads~\cite{curvature} to more specialized navigation for 
cyclists~\cite{komoot,bbbike,bike-citizens}, 
wheelchair routing~\cite{openrouteservice,wheelchair,crowdsourcing-mobility-impaired}, 
and routing according to sun or shade~\cite{shadow-as-route-2016,keithma2018parasol,deilami2020allowing}. 
None of the known solutions provide \emph{CCTV-aware} route planning 
and navigation, in particular focusing on privacy-first routing.

Olaverri Monreal et al.~\cite{shadow-as-route-2016} proposed a router that
follows shaded paths.
In 2018, Ma~\cite{keithma2018parasol} introduced \emph{Parasol}, which is a
routing solution that \textit{``uses high-resolution elevation data to simulate
sunshine and constructs routes that keep users in the sun or shade''}. The
author released the code as open-source~\cite{parasol2018github}.
Following the work of Ma~\cite{keithma2018parasol}, more recently Deilami et
al.~\cite{deilami2020allowing} introduced \emph{Shadeways}, a route planning
software to keep its users in shade. Their solution uses heat information and
tree shading data in the process of determining the route. 
While approaches such as \emph{Parasol}~\cite{keithma2018parasol} and
\emph{Shadeways}~\cite{deilami2020allowing} are inspiring and in theory could 
be adapted for CCTV-aware routing, they are not directly usable because of 
the differences in the underlying data requirements. 
%
In addition, neither \emph{Parasol} nor \emph{Shadeways} try to
solve some core challenges with OSM data. For example, in OSM a 
way object~\cite{osm-way} is represented as a line with no actual 
width on the map~\cite{osm-creaking}, whereas in reality a way is not a 
line but a space.

Some other authors use a safety-first approach when pursuing routing solutions. 
Bao et al.~\cite{safe-lighting} present a method for pedestrians to find safer
routes by investigating \textit{``the influences of landmark scarcity,
visibility, lighting condition, road width and turning tolerance on perceived
safety and comprehensiveness''}. 
Hirozaku et al.~\cite{street-illumination} also observe the illumination 
in order to provide a safer route for pedestrians. 
Keler and Mazimpaka~\cite{safety-aware-routing} aim towards routing in 
relatively dangerous areas within a city by utilizing 
Volunteered Geographic Information (VGI), i.e., crowdsourcing. 
Tessio et al.~\cite{customized-pedestrian-routes} provide a system which
enables users to define customized pedestrian routes with green areas, social
places and quieter streets.

Luxen and Vetter~\cite{osrm-first,luxen-vetter-2011} provide 
\emph{Open Source Routing Machine (OSRM)}, an even more general routing 
backend which enables users to write their own routing profiles in 
Lua code~\cite{osrm-profiles} as well as to use \emph{traffic updates} to set
weights between any two connected and routable nodes~\cite{osrm-traffic}. 
Using \emph{OSRM} profiles one can use any OSM tags to weigh the routing process.
In our work we use \emph{OSRM} and its routing profiles and \emph{traffic updates} 
as our routing solution.

In order to successfully manage and update the OSM data, relevant OSM 
data editors are required.
Indeed, many editors exist that can be used to edit OSM data. The most common 
of those with a graphical user interface are \emph{iD}, \emph{JOSM},
\emph{Potlach 2}, \emph{Vespucci} (Android) and \emph{Go Map!!}
(iOS)~\cite{osm-editors}. 
There are also more programmatic ways for editing OSM data. 
Two popular frameworks are \emph{Osmosis} (Java)~\cite{osmosis-osm-wiki} and 
\emph{Osmium} (C++)~\cite{osmium-osmcode}. \emph{Osmium} has also a Python
wrapper, \emph{PyOsmium}~\cite{pyosmium-osmcode} which is used in our
work. 
\emph{GeoPandas}~\cite{geopandas} and \emph{Shapely}~\cite{shapely}
are popular Python libraries for geometric manipulation of data, 
the latter of which is used in our work.

%


\section{Experimental setup}
\label{sec:exp}

In this section we present our experimental setup. 
We explore \emph{pedestrian routing} scenarios where the main constraint 
on the computations is about the ability of modelled cameras 
to \emph{recognize faces} or not. 
For \emph{privacy-first} option, the optimization is about routing the 
user in such a way that cameras are far away enough in order for the
\emph{face recognition} task to be successful.
For \emph{safety-first} option, the optimization is about routing the 
user in such a way that user is at anytime close enough to at least one 
camera which can perform \emph{face recognition} task successfully. 
The routing experiments are run on data collected from and relevant to 
the city of Jyv\"askyl\"a, Finland. 
In the future, we plan to extend the experiments to a more global scale.

\subsection{Short introduction to routing engine}
\label{sec:exp-osrm}

We base our experimental routing setup on \emph{Open Source Routing Machine
(OSRM)}~\cite{luxen-vetter-2011} and its routing profiles~\cite{osrm-profiles}
as well as its \emph{traffic updates}~\cite{osrm-traffic}. However, the main
contribution from a routing perspective relies on how we process OSM XML data,
which is essentially an XML file containing a list of certain OSM elements:
nodes, ways, and relations~\cite{osm-xml}. Data manipulation is needed to
mitigate the fact that an OSM way object is primarily a line without
width~\cite{osm-way}. In order to reach sub-meter routing accuracy required 
for \emph{CCTV-aware} (e.g., 1 meter can mean a difference between 
face recognition and face detection ability for the camera),
the \emph{OSM ways} need to have relevant width information associated. 
This information is required so that it is possible to allow routing through 
a way if only the other side of it is under the scope of a CCTV camera.

\begin{figure}[htb!]
  \centering
  \includegraphics[width=\columnwidth]{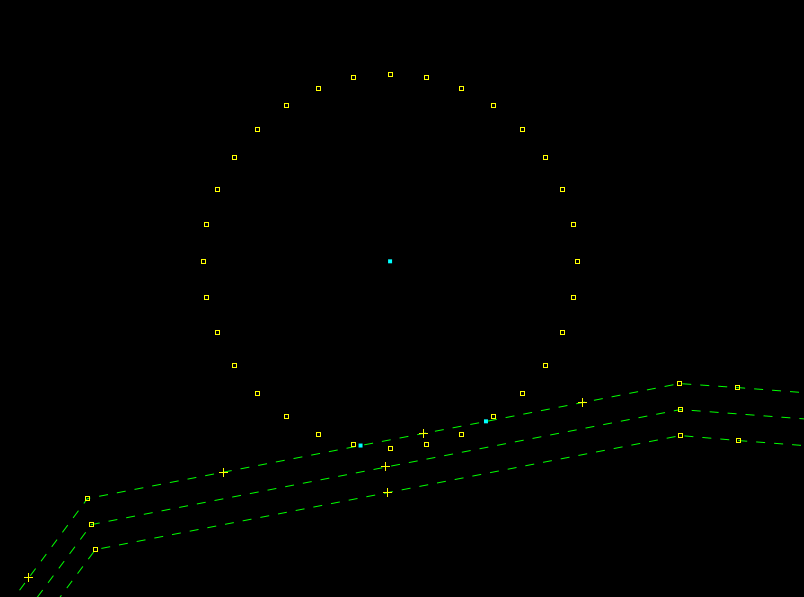}
    \caption{Sample from our OSRM engine: field of ``coverage'' (e.g., could be: face detection, face recognition - see Section~\ref{sec:exp-CCTV-cover}) 
    of a CCTV camera modelled as 360-degree view intersecting the edge of an OSM way that we then artificially split into two.}
  \label{fig:split_way}
\end{figure}

To mitigate the limitation of \emph{OSM way} objects having no width~\cite{osm-way},
Lucas-Smith proposes representing streets more realistically as boxes or areas
instead of lines~\cite{osm-creaking}. This has inspired the approach used in
our work. Instead of handling all the ways in an OSM file, we modify OSM data
locally by splitting a way into two whenever there is a CCTV camera nearby. Then,
if the scope of the CCTV camera covers only some of the way splits, it is still
possible to traverse the others while maintaining privacy. In a screenshot of a
view in \emph{JOSM}~\cite{josm,josm-wiki} (see Figure~\ref{fig:split_way}), the
teal-colored nodes (cyan squares) represent the camera and the points where one would enter under the
scope of the camera. As the scope does not cover the whole way, namely all of
the way splits, traversing the way still preserves privacy. 
It may however required that for the example in Figure~\ref{fig:split_way}, 
the user is explicitly instructed to walk at the opposite edge of the street 
when arriving close to the area of CCTV camera coverage. 
Such granularity is challenging in many ways, including GPS accuracy, 
and even though it is not implemented in our routing system at this moment
we plan it in our future roadmap. 

\subsection{Mapping of CCTV cameras}
\label{sec:exp-CCTV-map}

Before starting our experiments on different CCTV camera coverage models and 
corresponding route calculations, we have mapped a large number of CCTV 
cameras around Jyv\"askyl\"a, which is a city in Central Finland with an 
urban area of about 100 square kilometers and a population of about 138.000.
We mainly focused on its downtown and busiest areas.

\begin{figure}[htb!]
  \centering
  \includegraphics[width=\columnwidth]{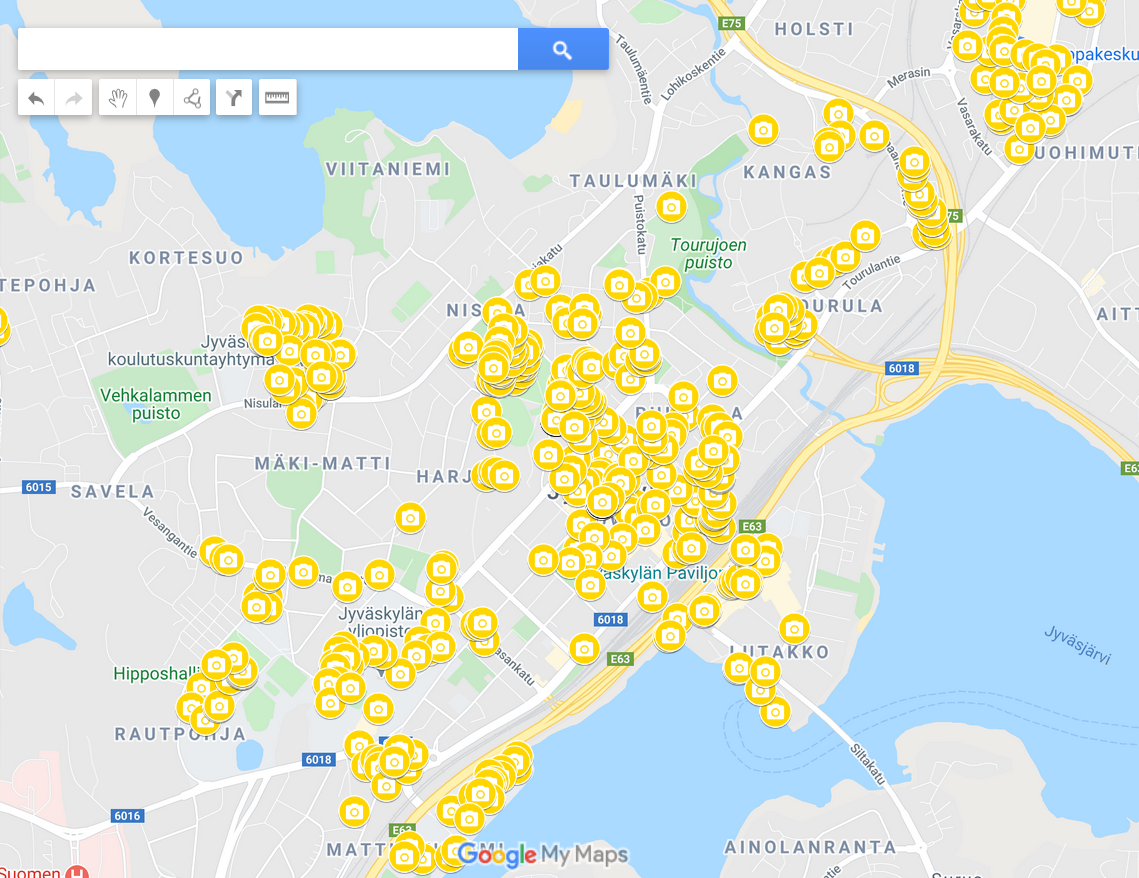}
  \caption{Data sample depicting a large portion of the entire set of 
  about 450 CCTV cameras that we mapped in Jyv\"askyl\"a.}
  \label{fig:img_cam}
\end{figure}

\begin{figure}[htb!]
  \centering
  \includegraphics[width=\columnwidth]{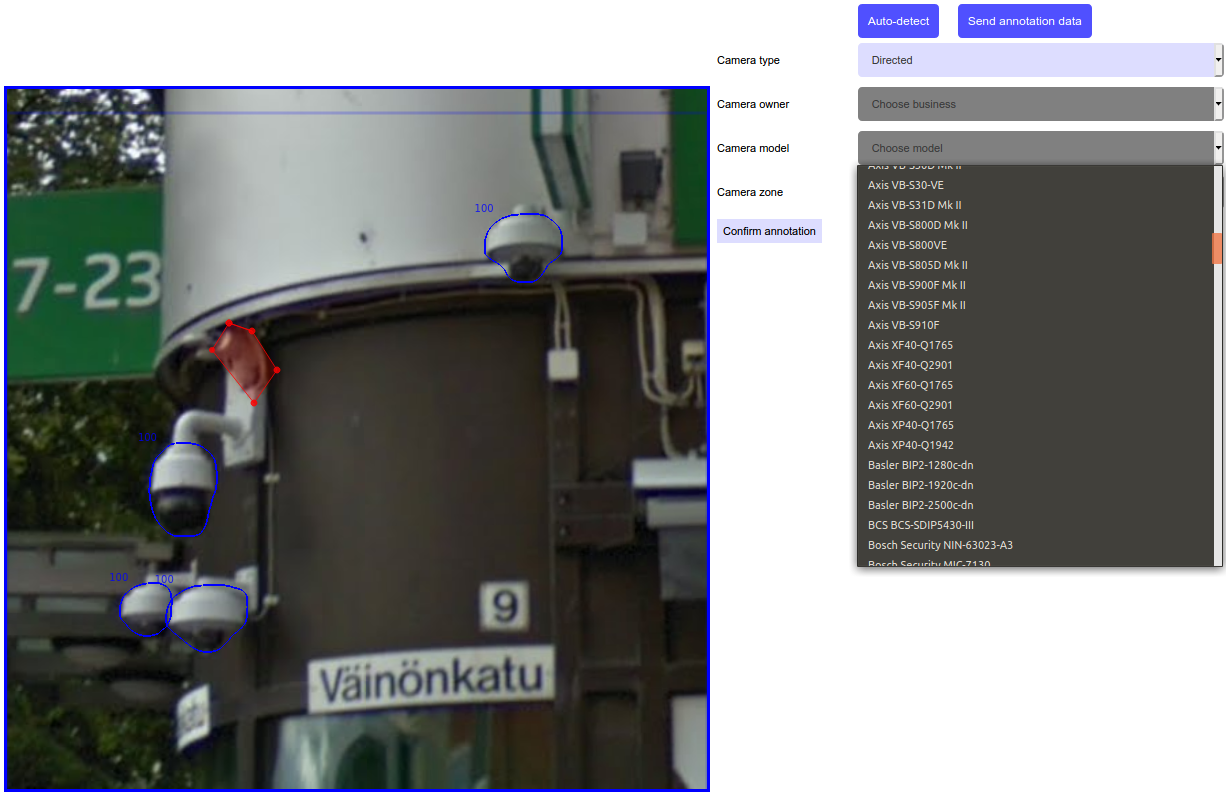}
  \caption{Screenshot from working with our novel and unique annotation, 
  autodetection and crowdsourcing tool for Computer Vision running as 
  \emph{in-the-browser extension} -- in \textcolor{blue}{blue are \emph{round} cameras} 
  autodetected with pixel-accuracy by the ML models; in \textcolor{red}{red are \emph{directed} cameras} 
  manually annotated by the user.}
  \label{fig:img_plugin}
\end{figure}

We have managed to map around 450 cameras in the 
city of Jyv\"askyl\"a, as partly depicted in Figure~\ref{fig:img_cam}. 
For now, their localization and mapping was mostly performed manually 
(about 400 cameras) by moving around the city to discover CCTV/surveillance cameras. 
The rest (about 50 cameras) were localized and mapped using a novel and 
unique \emph{in-browser plugin for annotation and crowdsourcing} tool (Figure~\ref{fig:img_plugin}) 
that we have specifically developed for this project and its future needs. 
Thus, our browser annotation tool allows us to use large-scale crowdsourcing 
for computer vision training purposes. It also enables us to annotate and 
to precisely localize CCTV cameras directly from the street level 
imagery. 
As the annotation tool is a plugin running within the browser, it is also 
perfectly positioned to parse the browser URLs. Therefore this allows us 
whenever possible and available to extract the geo-location information from 
online services such as Google Street View, and then link it precisely to 
the annotated street-level imagery. 

\subsection{Coverage of CCTV cameras}
\label{sec:exp-CCTV-cover}

Any given CCTV camera has more than a dozen properties that affect 
its performance and functionality. Even the same camera under different 
configuration (e.g., resolution, focal length) may have a totally different 
performance in exactly the same situation 
(e.g., may be able to recognize a face or not). 
Moreover, different CCTV cameras have different technical properties and 
operational settings. 
When it comes to qualitative tasks, such as license plate identification 
or face recognition, there must be a way to compare CCTV cameras regardless 
of their slightly different capabilities and configurations.
Therefore, a metric such as Pixel Per Meter (PPM)~\footnote{For non-metric jurisdictions, Pixel Per Foot (PPF) is defined and used.} 
was introduced which is a measurement used to define the amount of 
potential image detail that a camera offers at a given distance. 
Subsequently, various video surveillance tasks (e.g., face recognition) 
have been assigned a minimum PPM that the camera must support in its 
operational environment (i.e., resolution, focal length, distance to object) 
in order to have a chance to successfully perform the said task. 
%
According to European Standard EN 62676-4:2015~\cite{en201562676}, the 
following PPM requirements are for various video surveillance tasks:
\begin{itemize}
\item Identification -- 250 PPM
\item Recognition -- 125 PPM
\item Observation -- 62 PPM
\item Detection -- 25 PPM
\item Monitoring -- 12 PPM
\end{itemize}
\emph{``Face recognition''} means that person can be recognized 
(but not necessarily identified) from the CCTV footage. 
\emph{``Automated License Plate Recognition (ALPR)''} means a distance 
at which cars plate is clearly seen and the car can be identified 
according to the CCTV footage. 
%
For example, for successful ALPR it is required a CCTV vs. object (car)
configuration of between 200 PPM (EU license plate) 
and 400 PPM (Middle East license plate). 
%

Using various experimental setups, the researchers and the manufacturing 
companies have concluded and confirmed that with CCTV surveillance grade cameras, 
on average, the persons can be detected and tracked at distances of 25--50m, 
and the faces can be recognized at distances of 15--20m~\cite{wheeler2010face,axis-id-rec}.

Therefore, when we model the mapped cameras, we assume the following:
\begin{itemize}
\item \emph{low-performance cameras} -- face recognition successful at a maximum of 10m from the camera
\item \emph{medium-performance cameras} -- face recognition successful at a maximum of 15m from the camera
\item \emph{high-performance cameras} -- face recognition successful at a maximum of 25m from the camera
\end{itemize}
These numbers (i.e., 10m, 15m, 25m) are subsequently driving the types 
of experiments we plan and test (see Section~\ref{sec:exp-CCTV-route}).

In this preliminary feasibility study, we also chose to employ coarse modelling and 
apply a set of over-simplifications that apply to a single experimental run, 
as follows:
\begin{itemize}
\item All cameras have a 360-degree Field of View (FoV)
\item All cameras have the same properties, i.e., have the same radius to which they have perform the 
\item All cameras up to the extent of ``coverage'' radius are able to successfully perform ``face recognition''
\end{itemize}
Indeed, under more realistic assumptions, the cameras would have varying 
``coverage'' radius, as well as part of them would be 360-degree view while 
others would have sectoral FoV. 
Also, the ``coverage'' radius to be used for modelling the cameras on 
the routes would vary according to camera settings 
(which are unknown to anyone except the camera owner and/or operator) 
and the surveillance task at hand (e.g., face recognition vs. person detection). 
We leave these improvements and refinements as future work.

\subsection{Route points selection}
\label{sec:exp-CCTV-route}

We choose city central area for our routing experiments. We also preferred 
the most popular walking routes as well as specific places were surveillance 
cameras can block passage or force routes to use another path. 
For example, there are railway paths crossing Jyv\"askyl\"a separating parts 
of downtown by very few bridges over the railway. Those bridges are well 
surveyed by CCTV, therefore it is expected that such railway bridge crossings 
would pose a challenge for privacy-minded users.
%
For example, the results for routing involving the bridges over the railway 
can be seen in Figures~\ref{fig:img_6s},~\ref{fig:img_6p10} and~\ref{fig:img_6p15}. 
The route selection decision was partly helped by the data-intelligence 
from e\"Alyteli Project, which runs a network of IoT sensors scattered 
across the city of Jyv\"askyl\"a and provides basic framework for a 
data-driven SmartCity future. The IoT sensor network can provide insights 
to possible routes that are most commonly used by pedestrians that carry 
mobile devices. 
The final selection of the route points (i.e., departure and arrival) used in 
our experiments are presented in Table~\ref{tab:routing-summary}. 

As already mentioned (Section~\ref{sec:intro}), our system supports 
\emph{privacy-first} and \emph{safety-first} routing options.
In our experiments we test and evaluate the following routing modes 
(see Section~\ref{sec:exp-CCTV-cover}, Table~\ref{tab:routing-summary}):
\begin{enumerate}
\item Safety mode with CCTV cameras modelled as 360-degree with 10m radius of ``coverage''
\item Privacy mode with CCTV cameras modelled as 360-degree with 10m radius of ``coverage''
\item Privacy mode with CCTV cameras modelled as 360-degree with 15m radius of ``coverage''
\item Privacy mode with CCTV cameras modelled as 360-degree with 25m radius of ``coverage''
\end{enumerate}
%


%
While in each of the present experiments we used fixed and same camera parameters 
(e.g., 360-degree, coverage radius), our system is now ready to support 
scenarios where each camera is different while having also non-360-degree 
coverage and a varying ``coverage'' radius (e.g., according to individual 
camera's model and settings). This will allow us to model more realistic 
scenarios as well as allow the users of our system to select what exactly 
they want safety for or privacy against -- face recognition, face identification, 
ALPR identification, etc. We leave such experiments as future work.


\section{Results}
\label{sec:results}

In this section we present the results of the routing experiments described earlier.
In Table~\ref{tab:routing-summary} we summarize those results (with the following notes 
~\footnote{\label{ftn:tab-rout:not-full} The \emph{route not complete} result is when some final distance cannot be included in the route (cameras would break the privacy-preserving requirement).}
~\footnote{\label{ftn:tab-rout:middle} The routes were intentionally instructed to have an intermediate stop-point at Kauppakatu 18 (62.241531, 25.745317).}
~\footnote{\label{ftn:tab-rout:no-route} The engine was unable to generate a privacy-preserving route close to arrival point (when using given cameras modelling and given OSM data).}
~\footnote{\label{ftn:tab-avg} The \emph{no route} results are ignored from the calculation of average data.}
). 
From the Table~\ref{tab:routing-summary} it is immediately obvious that a 
person who wants to preserve a higher level of privacy/anonymity is forced 
to choose a longer route -- on average between 1.5x and 5.0x longer, 
depending on the configurations of the CCTV cameras around the city.
While such a result was somehow intuitively expected, our results confirm such 
hypothesis in a data-driven and evidence-based manner. 

\begin{table*}[htb]
\centering
\caption{Summary of routing cases and experimental results for the downtown of Jyv\"askyl\"a, Finland.}
\label{tab:routing-summary}
  \begin{tabular}{lcccccc}
    \toprule

    Route ID & \specialcell{Departure point \\ (GPS)} & \specialcell{Arrival point \\ (GPS)} & Safety [10m] & \specialcell{Privacy [10m] \\ (overhead)} & \specialcell{Privacy [15m] \\ (overhead)} & \specialcell{Privacy [25m] \\ (overhead)} \\
    \midrule
    
    1 & \specialcell{Vapaudenkatu 34-36 \\ (62.240541, 25.747361)}    &   \specialcell{Satamakatu 3 \\ (62.237443, 25.756073)}  &  \specialcell{760m \\ Figure~\ref{fig:img_2s}}  &   \specialcell{1.1km (1.4x) \\ Figure~\ref{fig:img_2p10}}    &   \specialcell{1.3km (1.7x) \\ Figure~\ref{fig:img_2p15}}    &  \specialcell{3.9km (5.1x)  \\ Figure~\ref{fig:img_2p25}}   \\
    \midrule 
    
    2 & \specialcell{Vapaudenkatu 75 \\ (62.244948, 25.756137)}    &   \specialcell{Sep\"ankatu 4 \\ (62.242070, 25.752404)}  &  \specialcell{460m \\ Figure~\ref{fig:img_3s}}  &  \specialcell{480m (1.0x)  \\ Figure~\ref{fig:img_3p10}}   &   \specialcell{2.3km (5.0x)  \\ Figure~\ref{fig:img_3p15}}  &   \specialcell{no route (N/A)~\footnotemark[\getrefnumber{ftn:tab-rout:no-route}] \\ Figure~\ref{fig:img_3p25}}    \\
    \midrule
    
    3 & \specialcell{V\"ain\"onkatu 13 \\ (62.244868, 25.746953)}    &   \specialcell{V\"ain\"onkatu 7 \\ (62.243479, 25.750451)}  &   \specialcell{250m \\ Figure~\ref{fig:img_4s}}  &   \specialcell{810m (3.2x) \\ Figure~\ref{fig:img_4p10}}    &   \specialcell{880m (3.5x)~\footnotemark[\getrefnumber{ftn:tab-rout:not-full}] \\ Figure~\ref{fig:img_4p15}}    &  \specialcell{no route (N/A)~\footnotemark[\getrefnumber{ftn:tab-rout:no-route}] \\ Figure~\ref{fig:img_4p25}}    \\
    \midrule
    
    4 & \specialcell{Uno Savolan katu 26 \\ (62.238352, 25.758390)}    &   \specialcell{Hannikaisenkatu 29 \\ (62.240491, 25.751438)}  &  \specialcell{710m \\ Figure~\ref{fig:img_6s}}  &   \specialcell{1.8km (2.5x) \\ Figure~\ref{fig:img_6p10}}    &   \specialcell{3.3km (4.6x) \\ Figure~\ref{fig:img_6p15}}    &  \specialcell{4.8km (6.8x)   \\ Figure~\ref{fig:img_6p25}}    \\
    \midrule
    
    5 & \specialcell{Hannikaisenkatu 20 \\ (62.244678, 25.746803)}    &   \specialcell{V\"ain\"onkatu 13 \\ (62.242510, 25.755215)}  &  \specialcell{800m \\  Figure~\ref{fig:img_7s}}  &   \specialcell{790m (1.0x) \\  Figure~\ref{fig:img_7p15}}    &   \specialcell{790m (1.0x) \\  Figure~\ref{fig:img_7p15}}    &  \specialcell{1.2km (1.5x)~\footnotemark[\getrefnumber{ftn:tab-rout:not-full}]   \\  Figure~\ref{fig:img_7p25}} \\
    \midrule
    
    6 & \specialcell{Mattilanniemi 2 \\ (62.232165, 25.736117)}    &   \specialcell{Hannikaisenkatu 20 \\ (62.242510, 25.755215)}  &  \specialcell{2.2km~\footnotemark[\getrefnumber{ftn:tab-rout:middle}] \\  Figure~\ref{fig:img_8s}}  &   \specialcell{2.7km (1.2x)~\footnotemark[\getrefnumber{ftn:tab-rout:middle}] \\  Figure~\ref{fig:img_8p10}}    &   \specialcell{3.1km (1.4x)~\footnotemark[\getrefnumber{ftn:tab-rout:middle}] \\  Figure~\ref{fig:img_8p15}}    &  \specialcell{7.3km (3.3x)~\footnotemark[\getrefnumber{ftn:tab-rout:not-full}]~\footnotemark[\getrefnumber{ftn:tab-rout:middle}]   \\  Figure~\ref{fig:img_8p25}} \\
    \midrule

    \specialcell{--} & \specialcell{--} & \specialcell{Average}  &   860m  &   1.28km (1.5x)    &   1.95km (2.3x)    &   4.3km (5.0x)~\footnotemark[\getrefnumber{ftn:tab-avg}]   \\

	\bottomrule
\end{tabular}
\end{table*}

As Table~\ref{tab:routing-summary} shows, the increase factors for the 
privacy-preserving routes get higher when the cameras are assumed to be 
high-performance, i.e., the radius at which they can successfully perform 
\emph{face recognition} is assumed to be 25m or even 15m.
The safety-first routes resulted in an average distance of 860 meters, 
while privacy-first routes on average were 1.28km or 1.5x longer 
(camera ``coverage'' radius 10m), 1.95km or 2.3x longer (camera ``coverage'' radius 15m), 
and 4.3km or 5.0x longer (camera ``coverage'' radius 25m) respectively. 

\begin{figure}[htb]
  \centering
  \includegraphics[width=0.75\columnwidth]{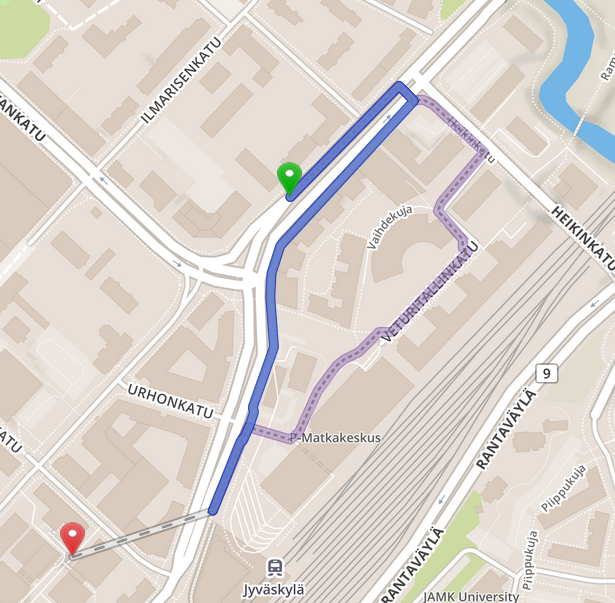}
  \caption{Route ID 2 -- privacy-first with 25m radius model -- no route.  \textsuperscript{\ref{ftn:tab-rout:no-route}}}
  \label{fig:img_3p25}
\end{figure}

\begin{figure}[htb]
  \centering
  \includegraphics[width=0.75\columnwidth]{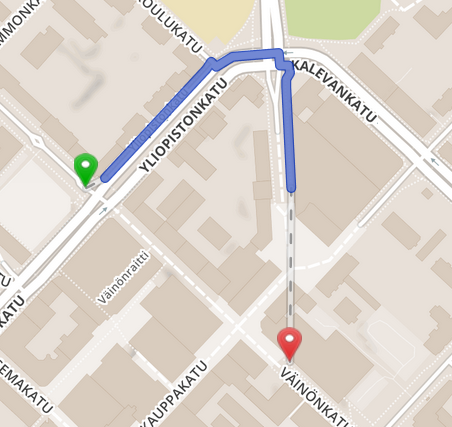}
  \caption{Route ID 3 -- privacy-first with 25m radius model --  no route. \textsuperscript{\ref{ftn:tab-rout:no-route}} }
  \label{fig:img_4p25}
\end{figure}

We also estimate that the increase factor of 5.0x for the privacy-first mode 
with camera ``coverage'' radius of 25m is a much lower lower bound than in 
reality.
One reason for such assumption is that the route engine was not able to draw 
point-to-point in some cases and we marked some routes when there were more 
than 100 meters missing from a fully expected route 
(e.g., Figures~\ref{fig:img_3p25} and ~\ref{fig:img_4p25}).
Another reason for such assumption is that route was avoiding cameras so far 
from the downtown that the route went into areas where cameras are not fully 
mapped into our database. If our database would contain 100\% of existing 
cameras, the privacy-first route with 25m ``coverage'' radius would result 
in even longer routes. 

Comparing routing results under different models provides some insights 
into what are the chances for person to avoiding cameras hence keeping 
their privacy/anonymity.
Our preliminary experiments and results underline the following observations. 
When modelling low-performance cameras, i.e., their maximum \emph{face recognition} 
``coverage'' radius is 10m, a person has quite decent chances of keeping 
their privacy/anonymity while navigating common paths and streets, even though 
the route lengths increases by a factor of 1.5x.
However, when modelling the routes under the assumption that the city employs 
medium- and high-performance cameras (i.e., ``coverage'' radius is 15m and 
25m respectively), it becomes increasingly problematic if not impossible to 
keep one's privacy/anonymity while moving around especially in the downtown area. 
For instance, the 15m ``coverage'' radius model there was one route that 
is marked as \emph{route not complete}, and the 25m ``coverage'' radius model 
there were two (2) \emph{route not complete} and two (2) \emph{no route} results. 
This also means that keeping the privacy/anonymity and avoiding cameras gets 
increasingly difficult when the ``coverage'' radius is increased from 15 to 25 meters.

\subsection{Analysis of some use-cases}
\label{sec:results-comp}

In this section we present and analyse a bit deeper some selected use-cases. 

\begin{figure}[htb!]
  \centering
  \includegraphics[width=0.75\columnwidth]{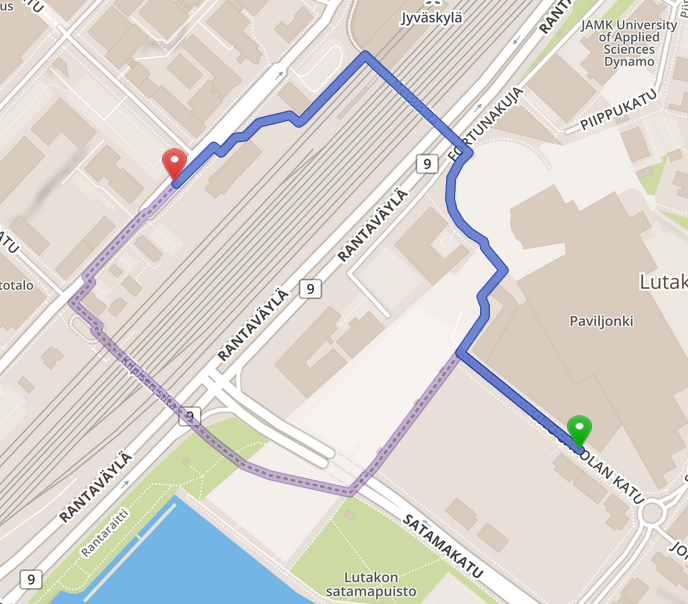}
  \caption{Route ID 4 -- safety-first with 10m radius model, distance 710m.}
  \label{fig:img_6s}
\end{figure}

\begin{figure}[htb!]
  \centering
  \includegraphics[width=0.75\columnwidth]{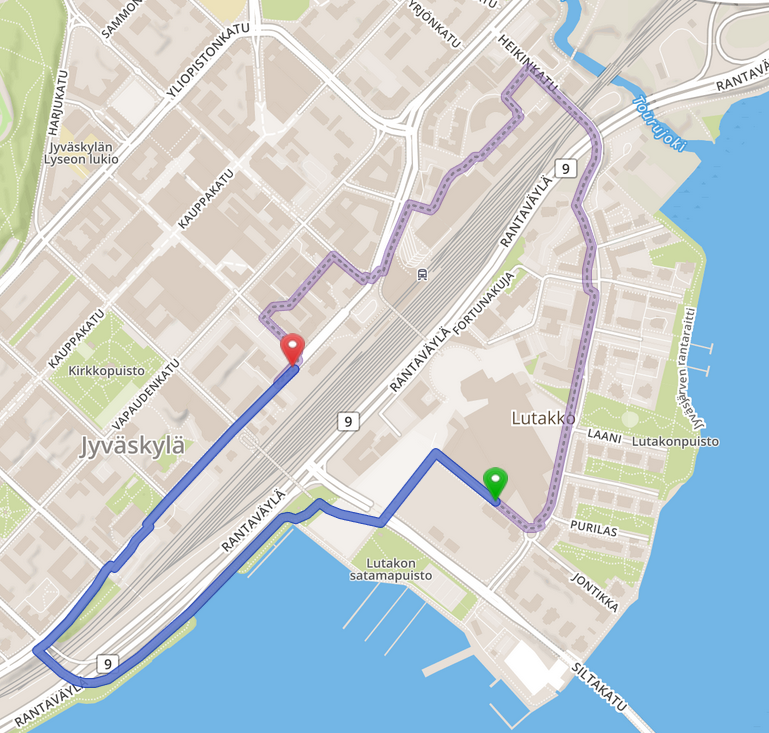}
  \caption{Route ID 4 -- privacy-first with 10m radius model, distance 1.8km.}
  \label{fig:img_6p10}
\end{figure}

\begin{figure}[htb!]
  \centering
  \includegraphics[width=0.75\columnwidth]{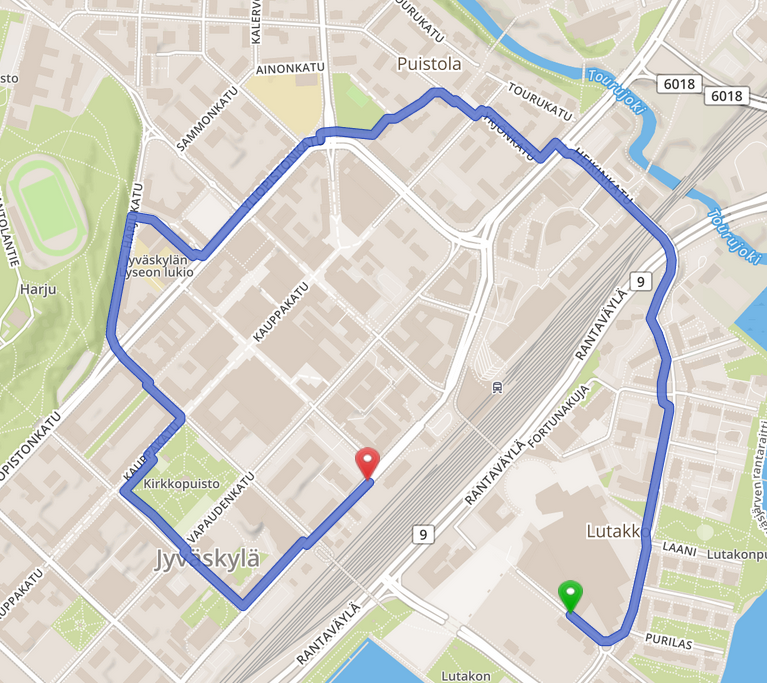}
  \caption{Route ID 4 -- privacy-first with 15m radius model, distance 3.3km.}
  \label{fig:img_6p15}
\end{figure}

One case is presented in the Figures~\ref{fig:img_6s} --~\ref{fig:img_6p15}.
It involves a railway crossing scenario, where several over-passage bridges 
and one under-passage path are the usual routes for pedestrians.
%
The \emph{safety-first} route (Figures~\ref{fig:img_6s}) uses one of the two 
nearest walk/cycle bridges to go over the railway. We confirmed these two 
bridges feature a set of CCTV cameras on their both ends, therefore it is 
virtually impossible to enter or exit one of those two bridges without 
entering a CCTV camera's \emph{face recognition} ``coverage''.
%
For the same departure and arrival points, the privacy-first route with a 
10m ``coverage'' model (Figure~\ref{fig:img_6p10}) is 2.5x longer, 
and with a 15m ``covered'' model (Figure~\ref{fig:img_6p15}) it is 4.6x longer.
We confirmed that in both privacy-first routes, both alternative bridge (15m model) 
and under-passage (10m model) are wider than the safety-first bridges hence 
provide more moving freedom, and do not have CCTV cameras immediately around 
them or at their entry/exit points. 
First, it can be observed that even at 15m ``coverage'' model, this causes 
pedestrian routing challenges in downtown area and forces a privacy-minded 
person to perform a huge loop around. Hence, increasing camera's 
\emph{face recognition} coverage with just 5m, causes a major change for 
privacy-first routings, which begs the important question -- 
\emph{Which should be the maximum privacy-invasion (e.g., face recognition) 
distances for CCTV cameras to operate at, so that they perform their safety 
and crime-deterrent function while preserving balance with privacy/anonymity?}
Second, this use-case clearly demonstrates that walk/cycle bridges with strict
CCTV camera coverage on its ends are quite challenging for privacy-minded persons. 
For example, for a person with reduced mobility who also is privacy-minded this 
is really a problem since they have to give-up on privacy because taking 5.0x 
longer route may not be an option for them.

\begin{figure}[htb!]
  \centering
  \includegraphics[width=0.75\columnwidth]{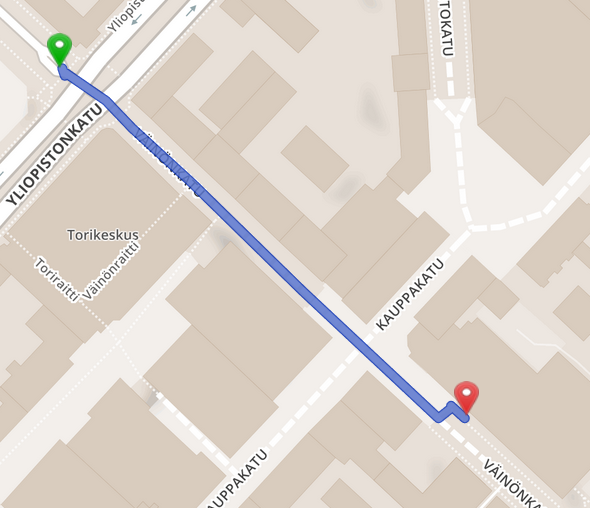}
  \caption{Route ID 3 -- safety-first with 10m radius model, distance 250m.}
  \label{fig:img_4s}
\end{figure}

\begin{figure}[htb!]
  \centering
  \includegraphics[width=0.75\columnwidth]{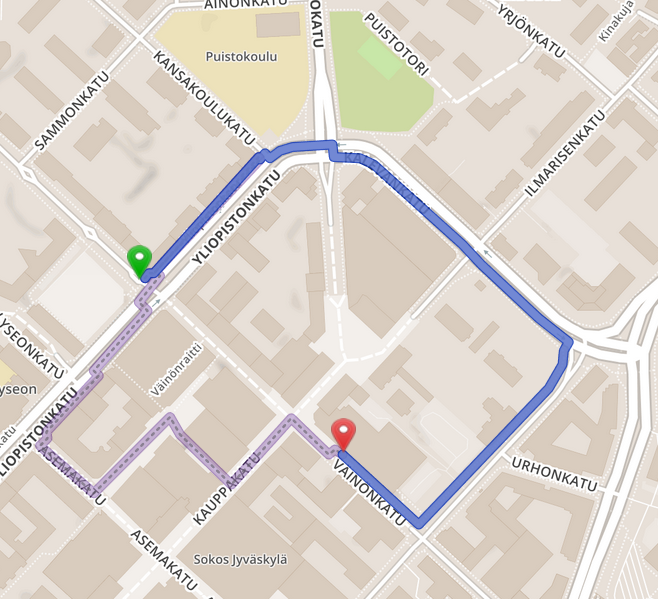}
  \caption{Route ID 3 -- privacy-first with 10m radius model, distance 810m.}
  \label{fig:img_4p10}
\end{figure}

\begin{figure}[htb!]
  \centering
  \includegraphics[width=0.75\columnwidth]{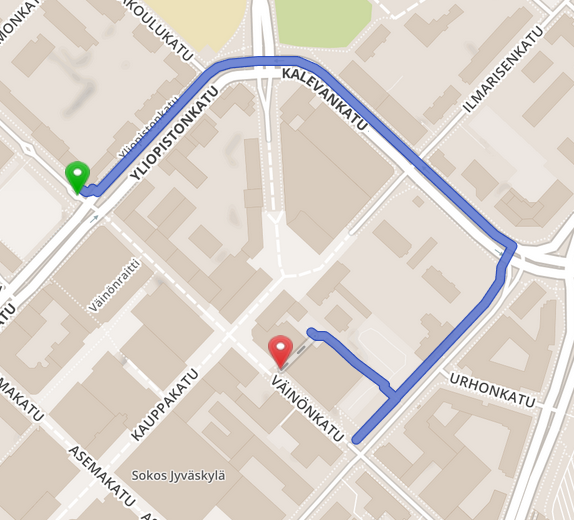}
  \caption{Route ID 3 -- privacy-first with 15m radius model, distance 880m. \textsuperscript{\ref{ftn:tab-rout:not-full}} }
  \label{fig:img_4p15}
\end{figure}

Another case is presented in the Figures~\ref{fig:img_4s}  --~\ref{fig:img_4p15}, 
and it covers a walking street in the center of Jyv\"askyl\"a (\emph{``V\"ain\"onkatu''}). 
In this case, the arrival point is located in one of the most camera-dense areas. 
The safety-first route (Figure~\ref{fig:img_4s}) takes straight line from departure to arrival points, 
both of which are located on V\"ain\"onkatu. 
While safety-first route goes straight, the privacy-first routes with both 
10m and 15m ``coverage'' models are forcing a loop around Kalevankatu. 
The main difference between the 10m model (Figure~\ref{fig:img_4p10}) and the 
15m model (Figure~\ref{fig:img_4p15}) is that the 15m model route does not 
reach the arrival point -- the route ends in the backyard of the commercial 
buildings which is the nearest point where routing can get while 
avoiding the en-route cameras.

\begin{figure}[htb]
  \centering
  \includegraphics[width=0.75\columnwidth]{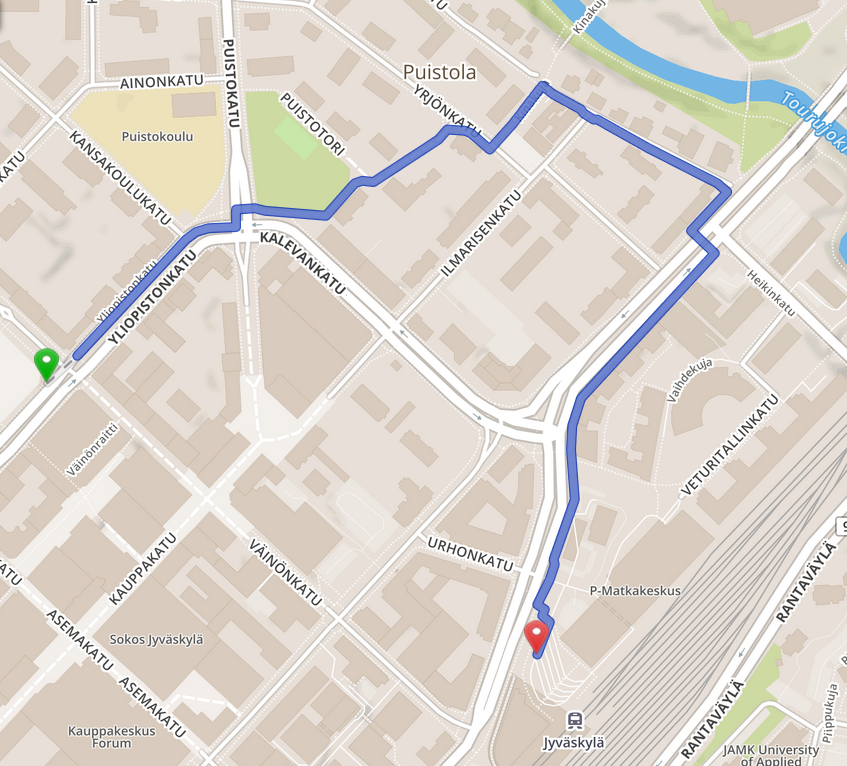}
  \caption{Route ID 5 -- privacy-first with 25m radius model, distance 1.2km. \textsuperscript{\ref{ftn:tab-rout:not-full}}}
  \label{fig:img_7p25}
\end{figure}

\begin{figure}[htb]
  \centering
  \includegraphics[width=0.75\columnwidth]{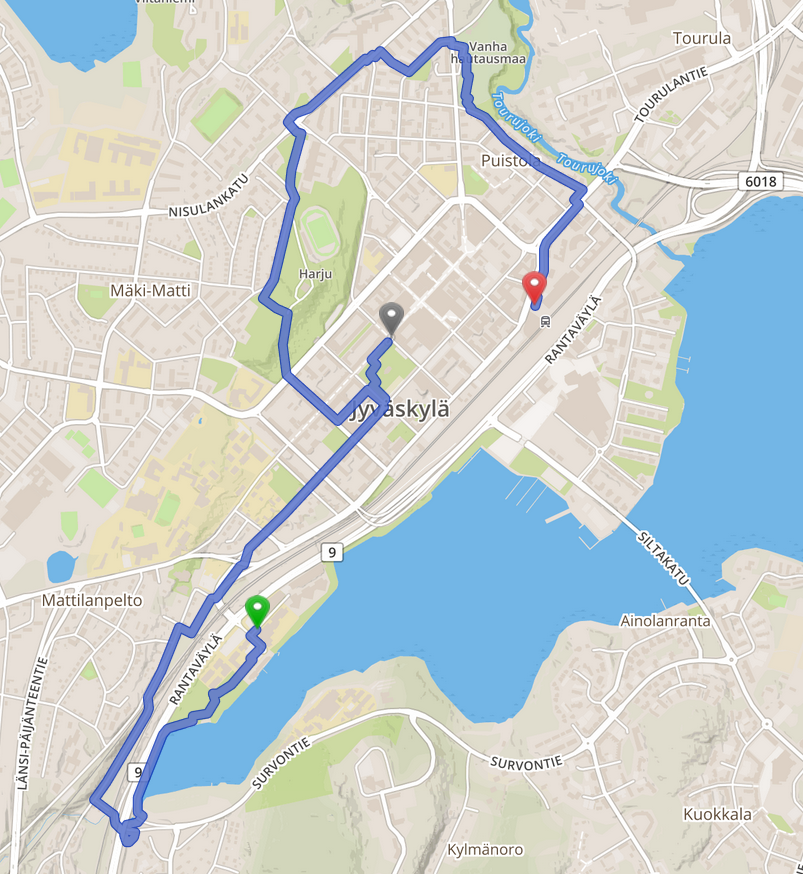}
  \caption{Route ID 6 -- privacy-first with 25m radius model, distance 7.3km. \textsuperscript{\ref{ftn:tab-rout:not-full}, \ref{ftn:tab-rout:middle}} }
  \label{fig:img_8p25}
\end{figure}

As yet another case, the Figures~\ref{fig:img_4p15},~\ref{fig:img_7p25} and ~\ref{fig:img_8p25} 
present situations where the route is not reaching from point-to-point. 
Figure~\ref{fig:img_8p25} presents an almost complete route, just missing 
about 10 meter from the ``middle point''. Figure~\ref{fig:img_7p25} shows 
that the route is missing departure point by about 10 meters. 
And in Figure~\ref{fig:img_4p15}, the final part of the V\"ain\"onkatu is 
blocked by the cameras and the route ends up in the backyard on the wrong 
side of the building.

It is also important to note the following final observations. 
Firstly, if the route cannot be calculated ``close enough'' to the departure 
or arrival point, the route is not output and not drawn. 
Secondly, the CCTV camera density is so high in the downtown area that it is not 
possible to compute privacy-preserving routes in some cases when ``coverage'' 
radius is over 20 meters, thus resulting in \emph{no route} situations. 
Thirdly, in some rare cases the \emph{safety-first} route can be longer than the 
\emph{privacy-first} one. For example, this can occur if both models 
(i.e., safety, privacy) are routed over the same rounded curve, 
and the cameras' ``coverage'' forces the privacy route towards the 
inner (shorter) curve while forcing the safety route towards the outer 
(longer) curve.

The rest of the visual results that correspond to our experiments can be 
found in Section~\ref{sec:apdx-samples}.

\section{Conclusion}
\label{sec:concl}

In this paper we studied the feasibility of developing CCTV-aware 
technology for privacy-first and safety-first routing and navigation, 
as well as the impact and difference between privacy-first and safety-first 
approaches.
For this, we created initial working prototypes of routing engine 
based on open-source OSRM framework. To support our initial exploration, 
we developed and experimented with somewhat simplified models involving 
CCTV cameras' mapping, cameras' coverage and corresponding privacy/safety impact. 
In addition, we relied in our efforts on some of our previous works and 
ideas such as accurate and instant detection and mapping of CCTV cameras 
using automated Computer Vision and manual crowdsourcing and annotation approaches. 

To prove our CCTV-aware approach is feasible, we run our experiments on real-world 
data-driven scenarios for downtown of the city of Jyv\"askyl\"a, Finland. 
Our results are encouraging (for further research) and 
worrying (for privacy-minded persons) at the same time.
Firstly, our experiments and results clearly demonstrate that our 
CCTV-aware technology works and can potentially bring lots of added-value to 
researchers, citizens, and policy-/decision-makers. 
This encourages us to improve and refine our models and implementations, 
and run much larger scale experiments based on richer and more accurate 
CCTV cameras datasets.
Therefore we invite interested parties 
(e.g., city administrations, governments) to contact us for joint-research 
towards SmartCities that are safe thanks to CCTV yet provide equal and 
privacy-preserving opportunities to all. 
Secondly, our experiments clearly show that there is an obvious and quantifiable 
impact for privacy-minded persons. 
The overall impact is negative both on quantitative dimension (e.g., distance increase), 
and qualitative dimension (e.g., some routes are impossible with full privacy/anonymity). 
For example, based on the data and models we used, the privacy-minded users 
suffer on average a distance increase between 1.5x to 5.0x (depending 
on the model applied) when compared to no-preference or safety-oriented users. 
While increasing and improving the datasets and the models will certainly 
result in different impact results, we expect the change to be 
even more towards a negative side for the privacy-minded users. 
One reason for such an assumption is because our current CCTV camera 
dataset is just a subset of all the CCTV cameras out there, i.e., the 
count of cameras we use in our modelling and route estimation is 
(considerably) smaller than in reality.

Finally, we argue that in order to move the discussion beyond subjective 
and futile debates such as ``CCTVs are good vs. CCTVs are bad'', 
technology must be efficiently used and data-driven decision and analysis must be applied.
This in turn prompts that the governments and administrations release and 
maintain as open-data detailed registries of both publicly- and 
privately-operated CCTV cameras. Such open-data will allow the users 
and decision makers to use CCTV-aware technology such as ours, 
therefore can make better decisions (e.g., routing for users, 
policies for decision-makers).

\section*{Acknowledgments}

We acknowledge grants of computer capacity from the 
Finnish Grid and Cloud Infrastructure (FGCI) 
(persistent identifier \texttt{urn:nbn:fi:research-infras-2016072533}).

Part of this research was kindly supported by the 
\emph{``17.06.2020 Decision of the Research Dean on research funding within the faculty''} 
grant from the Faculty of Information Technology of the University of Jyv\"askyl\"a, 
and the grant was facilitated and managed by Dr. Andrei Costin.

We also thank Prof. Timo H\"am\"al\"ainen and Riku Immonen for providing 
data-intelligence from e\"Alyteli Project~\cite{jyu-ealyteli} -- 
Ecological, intelligent and secured IoT services -- 
a project funded by the European Regional Development Fund (ERDF) and 
the partner companies involved. 

%




\bibliographystyle{IEEEtranS}
\bibliography{paper,paper-lauri}


\section*{Appendix}

\subsection{All routing samples}
\label{sec:apdx-samples}

\begin{figure}[htb!]
  \centering
  \includegraphics[width=0.75\columnwidth]{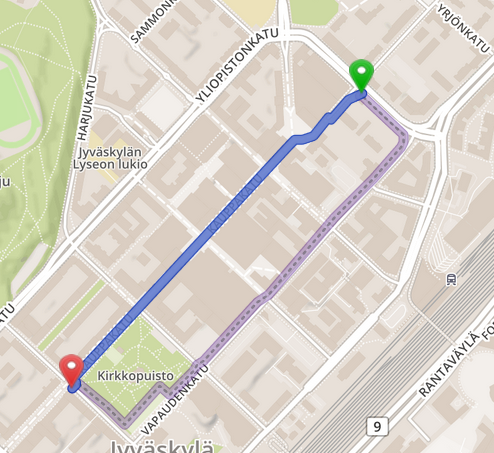}
  \caption{Route ID 1 -- safety-first with 10m radius model, distance 760m.}
  \label{fig:img_2s}
\end{figure}

\begin{figure}[htb!]
  \centering
  \includegraphics[width=0.75\columnwidth]{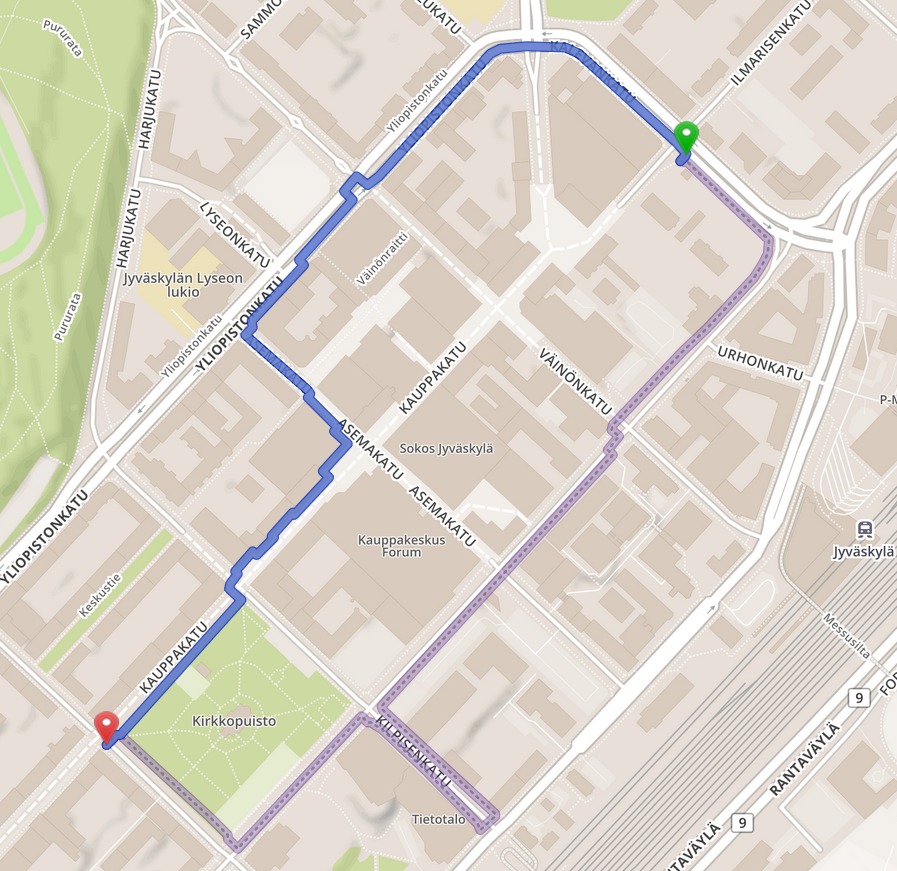}
  \caption{Route ID 1 -- privacy-first with 10m radius model, distance 1.1km.}
  \label{fig:img_2p10}
\end{figure}

\begin{figure}[htb!]
  \centering
  \includegraphics[width=0.75\columnwidth]{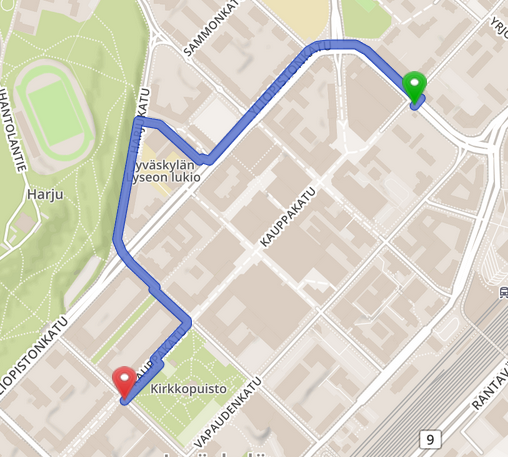}
  \caption{Route ID 1 -- privacy-first with 15m radius model, distance 1.3km.}
  \label{fig:img_2p15}
\end{figure}

\begin{figure}[htb]
  \centering
  \includegraphics[width=0.75\columnwidth]{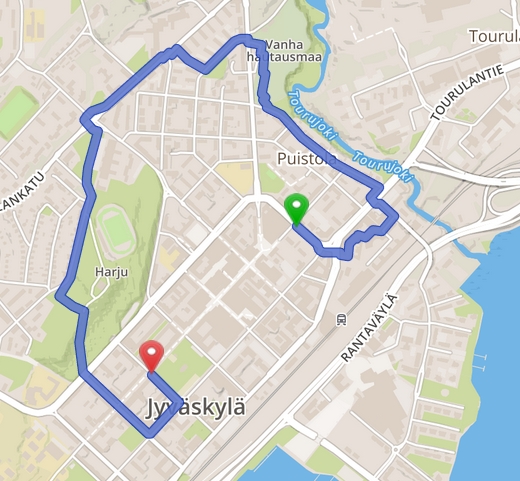}
  \caption{Route ID 1 -- privacy-first with 25m radius model, distance 3.9km.  \textsuperscript{\ref{ftn:tab-rout:no-route}}}
  \label{fig:img_2p25}
\end{figure}

\begin{figure}[htb]
  \centering
  \includegraphics[width=0.75\columnwidth]{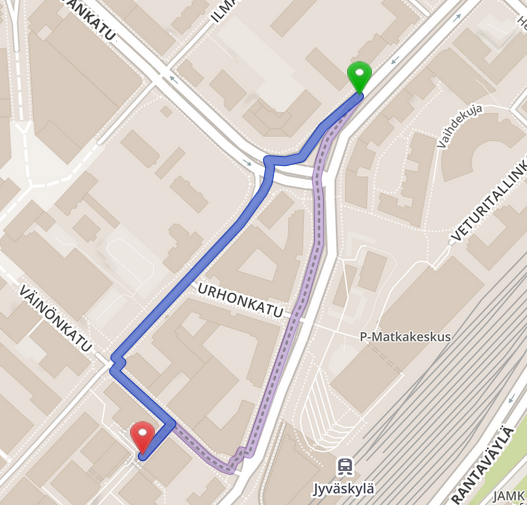}
  \caption{Route ID 2 -- safety-first with 10m radius model, distance 460m.}
  \label{fig:img_3s}
\end{figure}

\begin{figure}[htb]
  \centering
  \includegraphics[width=0.75\columnwidth]{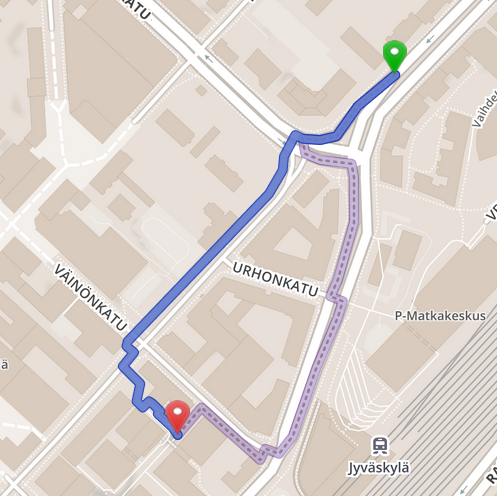}
  \caption{Route ID 2 -- privacy-first with 10m radius model, distance 480m.}
  \label{fig:img_3p10}
\end{figure}

\begin{figure}[htb]
  \centering
  \includegraphics[width=0.75\columnwidth]{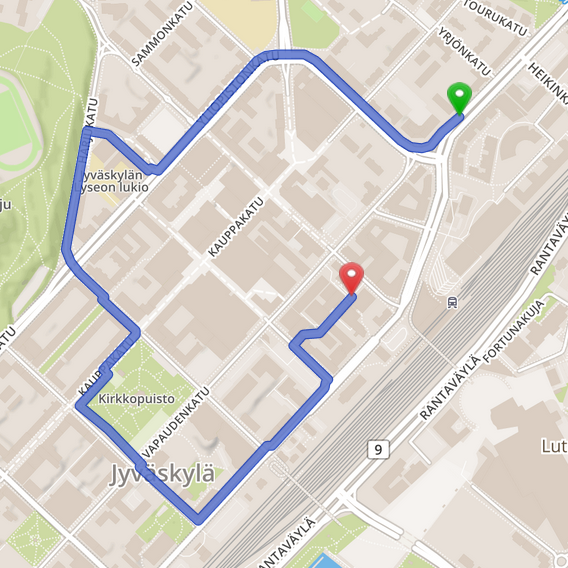}
  \caption{Route ID 2 -- privacy-first with 15m radius model, distance 2.3km.}
  \label{fig:img_3p15}
\end{figure}

\begin{figure}[htb]
  \centering
  \includegraphics[width=0.75\columnwidth]{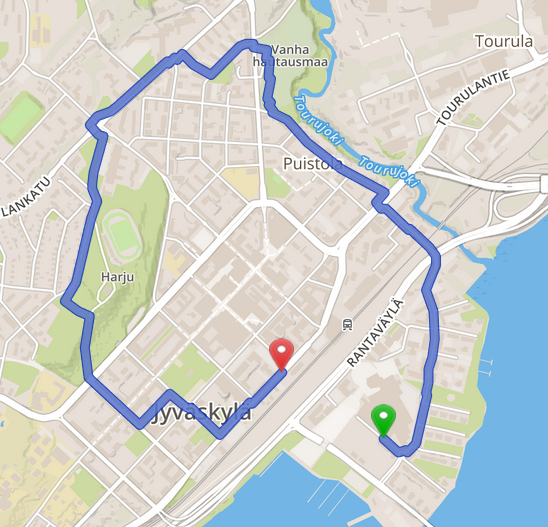}
  \caption{Route ID 4 -- privacy-first with 25m radius model --  distance 4.8km.}
  \label{fig:img_6p25}
\end{figure}

\begin{figure}[htb]
  \centering
  \includegraphics[width=0.75\columnwidth]{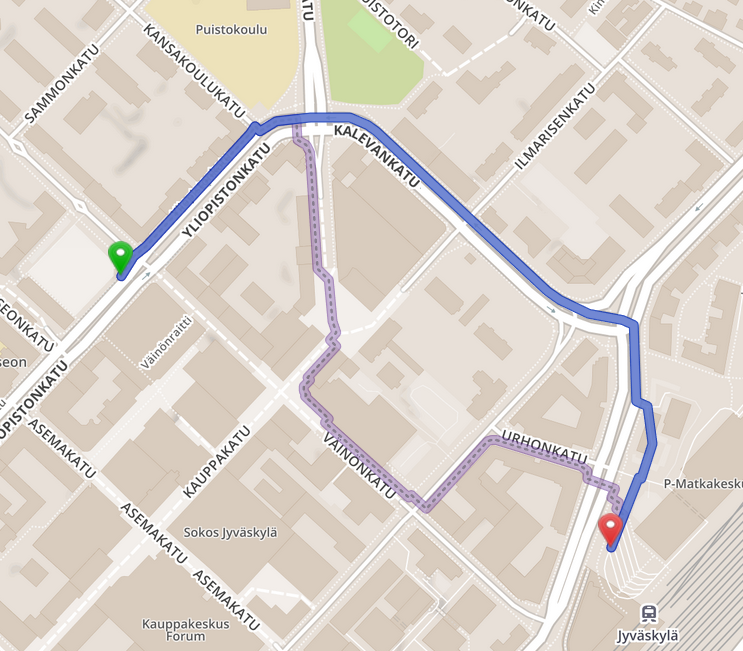}
  \caption{Route ID 5 -- safety-first with 10m radius model, distance 800m.}
  \label{fig:img_7s}
\end{figure}

\begin{figure}[htb]
  \centering
  \includegraphics[width=0.75\columnwidth]{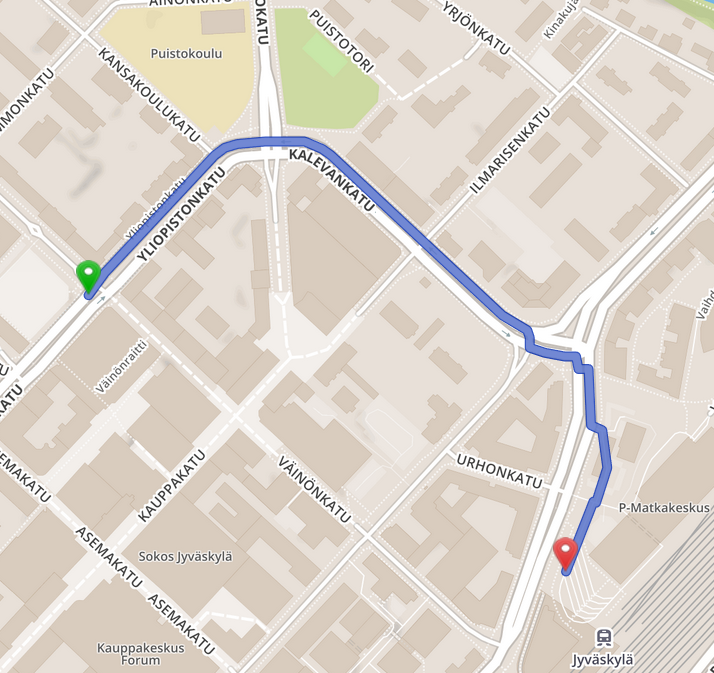}
  \caption{Route ID 5 -- privacy-first with 10 and 15m radius model, distance 790m.}
  \label{fig:img_7p15}
\end{figure}

\begin{figure}[htb]
  \centering
  \includegraphics[width=0.75\columnwidth]{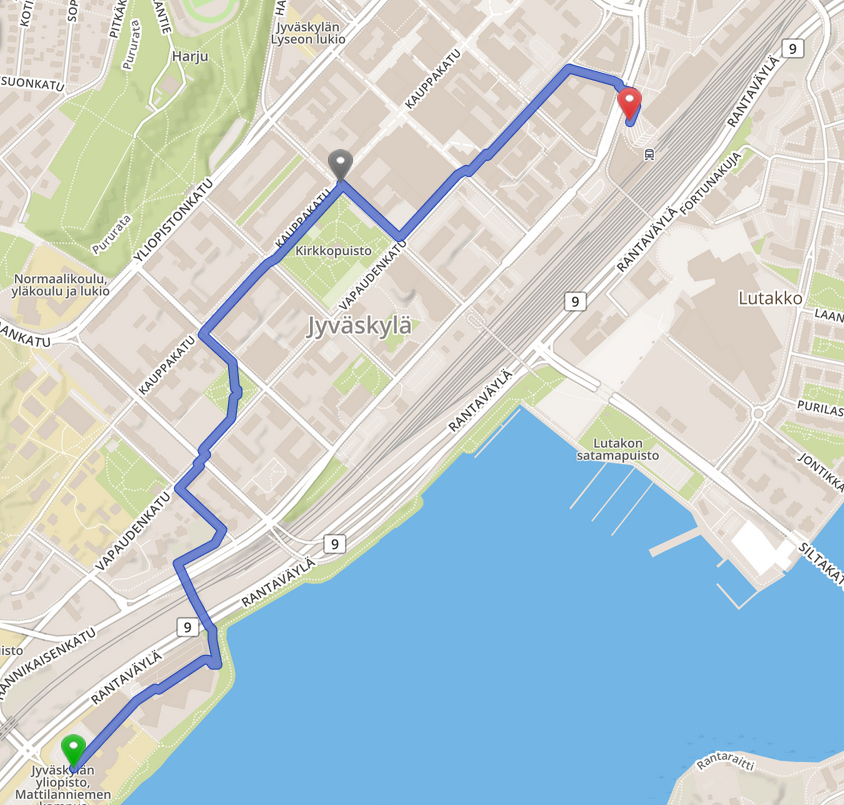}
  \caption{Route ID 6 -- safety-first with 10m radius model, distance 2.2km. \textsuperscript{\ref{ftn:tab-rout:middle}} }
  \label{fig:img_8s}
\end{figure}

\begin{figure}[htb]
  \centering
  \includegraphics[width=0.75\columnwidth]{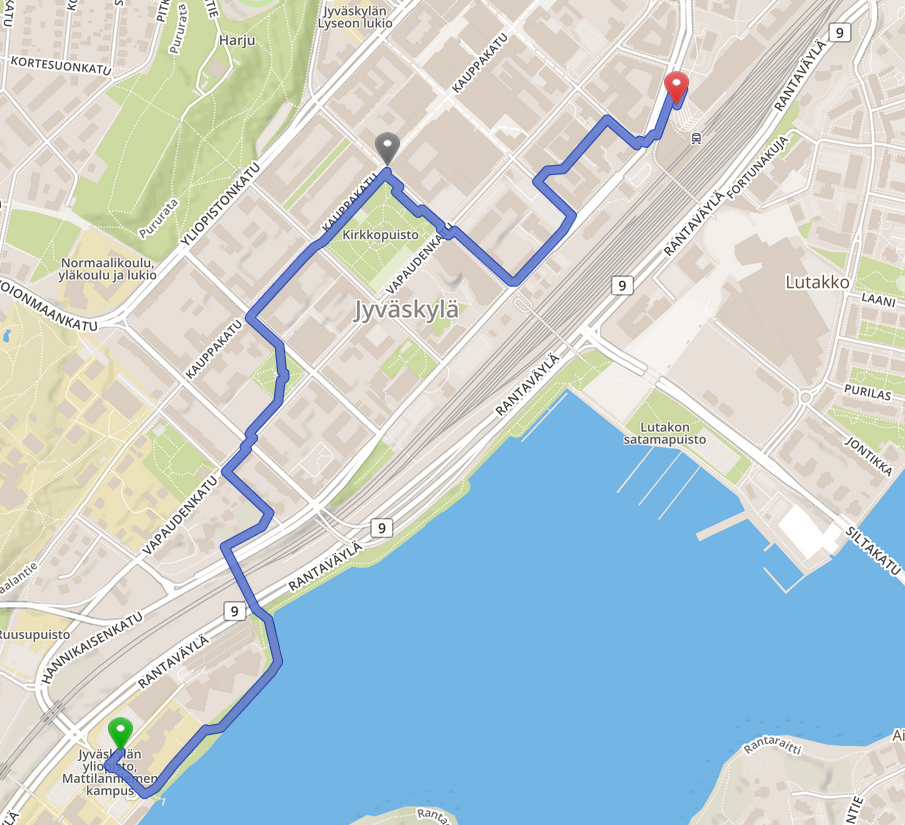}
  \caption{Route ID 6 -- privacy-first with 10m radius model, distance 2.7km. \textsuperscript{\ref{ftn:tab-rout:middle}} }
  \label{fig:img_8p10}
\end{figure}

\begin{figure}[htb]
  \centering
  \includegraphics[width=0.75\columnwidth]{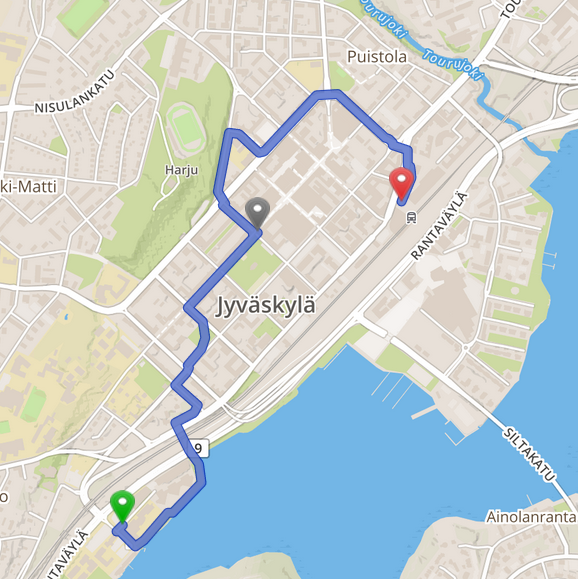}
  \caption{Route ID 6 -- privacy-first with 15m radius model, distance 3.1km. \textsuperscript{\ref{ftn:tab-rout:middle}} }
  \label{fig:img_8p15}
\end{figure}

\end{document}